# Spin Orbit Torque on a Curved Surface

Seng Ghee Tan[1†], Che Chun Huang[2], Mansoor B. A. Jalil[3], Zhuobin Siu[3]

(1) Department of Optoelectric Physics, Chinese Culture University, 55 Hwa-Kang Road, Yang-Ming-Shan, Taipei 11114, Taiwan

(2) Department of Physics, National Taiwan University, Taipei 10617, Taiwan

(3) Department of Electrical and Computer Engineering, National University of Singapore, 4 Engineering Drive 3, Singapore 117576

ABSTRACTS

We provide a general formulation of the spin-orbit coupling on a 2D curved surface. Considering the wide applicability of spin-orbit effect in spinor-based condensed matter physics, a general spin-orbit formulation could aid the study of spintronics, Dirac graphene, topological systems, and quantum information on curved surfaces. Particular attention is then devoted to the development of an important spin-orbit quantity known as the spin-orbit torque. As devices trend smaller in dimension, the physics of local geometries on spin-orbit torque, hence spin and magnetic dynamics shall not be neglected. We derived the general expression of a spin-orbit anisotropy field for the curved surfaces and provided explicit solutions in the special contexts of the spherical, cylindrical and flat coordinates. Our expressions allow spin-orbit anisotropy fields and hence spin-orbit torque to be computed over the entire surfaces of devices of any geometry.

Corresponding author:
† Seng Ghee Tan (Prof)
Department of Optoelectric Physics,
Chinese Culture University,
55, Hwa-Kang Road, Yang-Ming-Shan,
Taipei, Taiwan 11114 ROC
(Tel: 886-02-2861-0511, DID:25221)

† Email: csy16@ulive.pccu.edu.tw ; tansengghee@gmail.com

PACS:

2## INTRODUCTION

The rise of spin physics in condensed matter and nanoscience is so prolific that spinor-based expressions is now a hallmark formulation in nearly all the modern research fields like the spintronics **[1-5]**, graphene **[6,7]**, topological insulators **[8,9]**, topological Weyl and Dirac systems **[10, 11]**, atomic and optical physics **[12]**, quantum computation and information science **[13, 14]**. Central to the success of spin in science is the advancements in our understanding of the spin-orbit coupling. On the level of pure science, it is a Dirac relativistic phenomenon that exists in the non-relativistic limit. From the condensed matter point of view, spin-orbit coupling is a highly accessible physics that exists in bulk and structural forms in materials like the semiconductor, semimetal, as well as heterostructures comprising these materials. In applied science, it is simply an effective magnetic field that can be controlled via electrical means to perform electronic functions like in transistors and memory. In this article, we would focus on the spin orbit coupling in a specific area known as the spin-orbit torque. In magnetic memory, spin current is normally injected into magnetic materials to flip the magnetic moments via the force physics of spin transfer torque **[15]**. In nanoscience systems, spin torque is usually studied in the ferromagnetic materials, in which the net magnetization, reacts to the injection or the "creation" of an accumulation of spin moments in the material. One example of such "creation" is via the spin-orbit effect. With spin-orbit effect identified for the new role, the force involved in this process would be known as the spin-orbit torque **[16-22]**.

This paper is dedicated to providing a general formulation of the spin-orbit coupling and hence the spin-orbit torque in a 2D environment that comprises curved surfaces, with the 2D planar surfaces being a special case of the curves. Such formulation is important because as device becomes smaller, the physics of local geometries on spin-orbit coupling and its torque shall not be neglected. In fact, as spin-orbit coupling is versatile in many modern fields, a general formulation is simply right for the occasion. For example, curve physics has recently been studied in Dirac-electronic transport due to the topological surface states **[23-26]**. Curves were studied for their effects on inducing topological transitions **[27]** in a Rashba system, as well as shifting the band-inversion point **[28]** in the BHZ topological-insulator. In spintronics, the effect of curve on spin current **[29-31]** spin Hall **[32]**, and spin Chern number **[33]** have also been studied.

In this paper, we present a general formulation of the spin-orbit torque in a 2D condensed matter system. We will begin with a 3D infinitesimal bulk in which a 2D surface can be retrieved later by setting $q_3 \to 0$. Let the electric field penetrating the bulk be normal to the surface (to be denoted by $\boldsymbol{e_3}$) as shown in Fig.1 below. The spin-orbit energy of a charged particle contained therein would be given by

$$H_{soc} = \frac{\alpha^3}{\hbar} \boldsymbol{\sigma} \cdot (\boldsymbol{P} \times \boldsymbol{e_3})$$

(1)

where $\boldsymbol{\sigma}$ is the Pauli matrices, and $\alpha^3$ is a constant to characterize the spin-orbit strength. While Equation (1) is in the general form, constant $\alpha^3$ would reflect the material property and the type of spin-orbit coupling, e.g. Rashba, Dresselhaus, in bulk, heterostructures or topological surface states. The momenta and Pauli matrices on the 2D surface take on coordinates $\boldsymbol{e_1}, \boldsymbol{e_2}$. And they are related to the electric field direction by $\boldsymbol{e_1} \times \boldsymbol{e_2} = |n_3| \boldsymbol{e_3}$. It can then be shown that $|n_3| = \frac{\sqrt{G}}{G_{33}}$, which leads to

$$H_{soc} = \left(\frac{\alpha^3}{\hbar}\right) \sigma^v \boldsymbol{e_v} \cdot \left[P^k \boldsymbol{e_k} \times (\boldsymbol{e_1} \times \boldsymbol{e_2}) \frac{G_{33}}{\sqrt{G}}\right]$$

(2)

where $(v, k)$ runs over coordinates $(q_1, q_2)$. $G_{33} = \frac{\partial \boldsymbol{R}}{\partial q_3} \cdot \frac{\partial \boldsymbol{R}}{\partial q_3}$ is the 3D metric of the system and $G$ is the determinant given by $G = (G_{11}G_{22} - G_{12}G_{21})G_{33}$. Note that with $\alpha^3$ being contravariant, the upper and lower indices would balance out. The contravariant property of $\alpha^3$ will be discussed later.

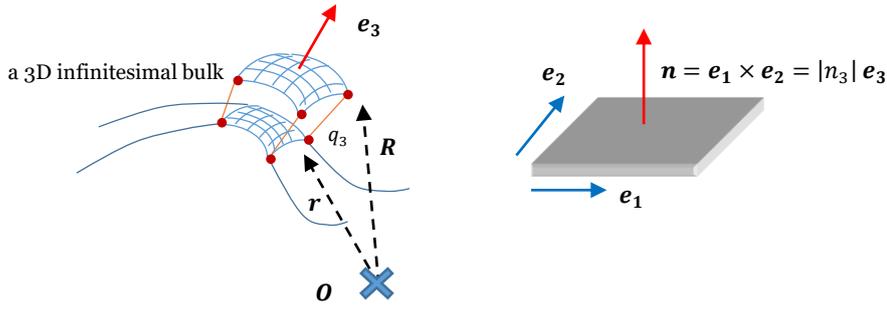

**FIG.1.** An infinitesimal bulk can be shrunk along $e_3$ to a 2D surface embeded in a 3D space. This is tantamont to taking the limit of $q_3 \to 0$.

Equation (2) is then written as follows

$$H_{SOC} = \left(\frac{\alpha^3}{\hbar}\right) P^k \sigma^v R_{kv3}$$

(3)

where $R_{kv3} = \frac{G_{33}}{\sqrt{G}}(G_{v1}G_{k2} - G_{v2}G_{k1})$. Expanding the above with the electric field pre-determined along $e_3$, discussion is kept to the 2D spin-orbit effect, and one has the general expression of

$$H_{soc} = \left(\frac{\alpha^3}{\hbar}\right)[\sigma^1 P^2 - \sigma^2 P^1]\sqrt{G}$$

(4)

Note that (1,2,3) is the abbreviated form of general coordinates $(q_1, q_2, q_3)$. A compact expression of the above is given by

$$H_{soc} = \frac{P^1}{m_e} S_1 + \frac{P^2}{m_e} S_2$$

(5)

where $S_1 = -\left(\frac{\alpha^3}{\hbar}\right) m_e \sqrt{G}\, \sigma^2$, $S_2 = \left(\frac{\alpha^3}{\hbar}\right) m_e \sqrt{G}\, \sigma^1$. Note that $(P^1, P^2)$ and $(\sigma^1, \sigma^2)$ can be found by transforming their Cartesian counterparts to the desired coordinates in a contravariant manner. When the general coordinates $(q_1, q_2, q_3)$ take on the Cartesian $(x, y, z)$, the term $\sqrt{G}$ goes to 1. The general Hamiltonian $H_{SOC}$ returns to the familiar spin-orbit system in the Cartesian frame,

$$H_{soc} = \left(\frac{\alpha^3}{\hbar}\right)(-P^x \sigma^y + P^y \sigma^x)$$

(6)

We will now examine the physical significance of $\alpha^3$, the spin-orbit constant that characterizes its strength. As $\alpha^3$ captures the material property pertaining directly to the strength of the electric field or the effective electric field in the case of the 2D spin-orbit effect e.g. the Rashba or the Dresellhaus, we will now explicitly express $\alpha^3$ in terms of its electric field as $\alpha^3 = \alpha' E^3$. Eq.(1) can now be rewritten to better reflect the actual scenario where the presence of the electric field is explicit

$$H_{soc} = \frac{\alpha^3}{\hbar} \boldsymbol{\sigma} \cdot (\boldsymbol{p} \times \boldsymbol{e_3}) \quad \to \quad H_{soc} = \frac{\alpha'}{\hbar} \boldsymbol{\sigma} \cdot (\boldsymbol{p} \times E^3 \boldsymbol{e_3})$$

(7)

Now $E^3$ would be the strength of the electric field normal to surface i.e. along $e_3$. For better clarity, we use the simple spherical surfaces for illustration in Fig.2 below.

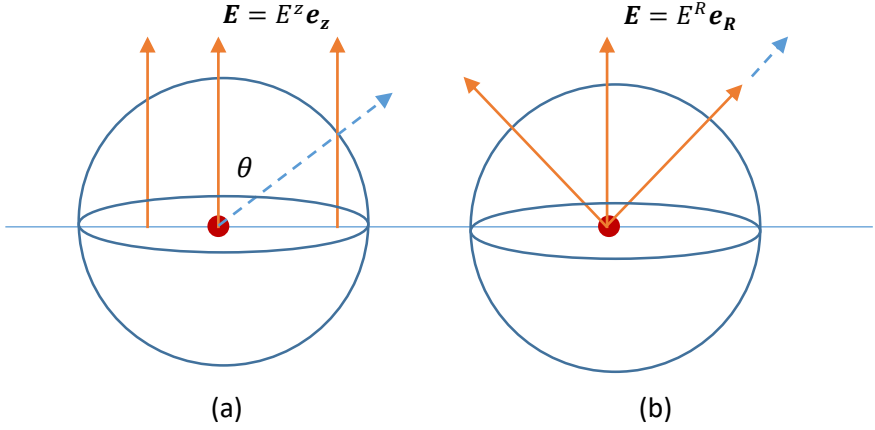

**FIG.2.** Illustration of the electric field orientations for the spherical surfaces. Solid lines are the actual electric fields, i.e. in (a) $\boldsymbol{E} = E^z \boldsymbol{e}_z$, in (b) $\boldsymbol{E} = E^R \boldsymbol{e}_R$. Dotted lines denote the orientation normal to the surface, i.e. $\boldsymbol{e}_3 = \boldsymbol{e}_R$.

Consider the case of a spherical surface now where $\boldsymbol{e}_3 = \boldsymbol{e}_R$. In the event that the actual electric field is $\boldsymbol{E} = E^z \boldsymbol{e}_z$, the expression for the electric field normal to the surface is $E^S = (E^z \boldsymbol{e}_z) \cdot \boldsymbol{e}^R$, and that would result in $E^S = E^z \cos\theta$. In the case where actual electric field is given by $\boldsymbol{E} = E^R \boldsymbol{e}_R$, one would obtain $E^S = (E^R \boldsymbol{e}_R) \cdot \boldsymbol{e}^R = E^R$. Therefore, in the event of a general surface marked by $\boldsymbol{e}_3$ or $\boldsymbol{e}^3$ and an electric field oriented along $\boldsymbol{e}_3$, the electric field normal to the surface is

$$E^S = (E^3 \boldsymbol{e}_3) \cdot \boldsymbol{e}^3 = E^3 \tag{8}$$

It thus becomes clear that $H_{soc} = \frac{\alpha^3}{\hbar} \boldsymbol{\sigma} \cdot (\boldsymbol{p} \times \boldsymbol{e}_3)$, with upper index 3, generalizes the representations for electric fields oriented vertical to a surface $\boldsymbol{e}_3$. The illustration above lends a clear physical meaning to the expressions of $S_1 = -\left(\frac{\alpha^3}{\hbar}\right) m_e \sqrt{G}\, \sigma^2$ and $S_2 = \left(\frac{\alpha^3}{\hbar}\right) m_e \sqrt{G}\, \sigma^1$. With $\alpha^3$ and hence $(S_1, S_2)$ properly understood, *a general formulation of the spin-orbit coupling on a 2D curved surface with electric field normal to the surface have thus been derived in Eq.(5).* In the event that the electric field is not restricted to the normal surface, one could derive using Eq.(2) and letting $(v, k)$ run over coordinates $(q_1, q_2, q_3)$ and obtain

$$S_1 = \left(\frac{\alpha^2}{\hbar}\right) m_e \sqrt{G}\, \sigma^3 - \left(\frac{\alpha^3}{\hbar}\right) m_e \sqrt{G}\, \sigma^2 \tag{9}$$

$$S_2 = \left(\frac{\alpha^3}{\hbar}\right) m_e \sqrt{G}\, \sigma^1 - \left(\frac{\alpha^1}{\hbar}\right) m_e \sqrt{G}\, \sigma^3 \tag{10}$$

$$S_3 = \left(\frac{\alpha^1}{\hbar}\right) m_e \sqrt{G}\, \sigma^2 - \left(\frac{\alpha^2}{\hbar}\right) m_e \sqrt{G}\, \sigma^1 \tag{11}$$

One should, however, not take for granted that in the event of $\alpha^2$, $\boldsymbol{e}_2$ would be normal to $\boldsymbol{e}_1$ and $\boldsymbol{e}_3$, and likewise for $\alpha^1$. The orthogonal property of the space metric continues to be enforced by $\boldsymbol{e}_1 \times \boldsymbol{e}_2 = \frac{\sqrt{G}}{G_{33}} \boldsymbol{e}_3$ throughout and the metric of the system is

$$\begin{bmatrix} G_{11} & G_{12} & G_{13} \\ G_{21} & G_{22} & G_{23} \\ G_{31} & G_{32} & G_{33} \end{bmatrix} = \begin{bmatrix} G_{11} & G_{12} & 0 \\ G_{21} & G_{22} & 0 \\ 0 & 0 & 1 \end{bmatrix} \tag{12}$$

The spin-orbit Hamiltonian would simply be

$$H_{soc} = \frac{P^1}{m_e}S_1 + \frac{P^2}{m_e}S_2 + \frac{P^3}{m_e}S_3$$

(13)

## SPIN-ORBIT TORQUE

In the context of the spin-orbit torque, the system under consideration would be a heterostructure that comprises the ferromagnet and the oxide layers (we cite the example of Ta\CoFeB\MgO **[21]**), with the ferromagnetic interface hosting a collective density of spin-orbit-induced spin moment denoted by $s$. At the same time, the ferromagnetic material possesses a collective density of intrinsic moment known as $m$. Therefore, the physics of spin-orbit torque arises due to the simultaneous presence of the kinetic energy, the spin-orbit energy and the magnetic energy in the system. The full Hamitonian, must now be presented to incorporate these energies as follows

$$H = \frac{1}{2m_e}P_a P^a + \frac{1}{2m_e}(P_a S^a + S^a P_a) + \sigma^a m_a$$

(14)

For generality, we let $a$ runs over $(q_1, q_2, q_3)$. In the above, the kinetic energy would be $\frac{1}{2m_e}P_a P^a = -\frac{i\hbar}{2m_e}\left(\partial_a + \frac{1}{2}\partial_a \ln G\right)P^a$. This is because when $P_a$ is quantized, the covariant property of the operator has to be accounted for. Now, a minimal coupling form is presented, which reflects intuitively the "forceful" physics of the Hamiltonian. Written in this way, $S_a$ relates directly to the non-Abelian gauge that has been studied as an origin of spin forces and phases in many systems **[5]**.

$$H = \frac{1}{2m_e}(P_a + S_a)(P^a + S^a) + G_{ab}\sigma^a m^b$$

(15)

Expanding the above,

$$H = -\frac{i\hbar}{2m_e}\left(\partial_a + \frac{1}{2}\partial_a \ln G\right)P^a \psi - \frac{i\hbar}{2m_e}\left(\partial_a + \frac{1}{2}(\partial_a \ln G)\right)S^a \psi + \frac{1}{2m_e}S^a P_a + G_{ab}\sigma^a m^b$$

(16)

One can now perform a local gauge transformation in the spin space such that the local frame would rotate to "some" axis – the choice of which would pretty much determine the physics to be revealed from those energies. The locally transformed system is given by

$$H' = -\frac{i\hbar}{2m_e}\left(\nabla_a + \frac{ie}{\hbar}A_a\right)(P^a + eA^a) - \frac{i\hbar}{2m_e}\left(\nabla_a + \frac{ie}{\hbar}A_a\right)US^a U^\dagger - US^a U^\dagger \frac{i\hbar}{2m_e}\left(\partial_a + \frac{ie}{\hbar}A_a\right) + G_{ab}U\sigma^a m^b U^\dagger$$

(17)

where $\nabla_a = \partial_a + \frac{1}{2}\partial_a \ln G \equiv \partial_a + \Gamma_a$ and $A_a = -i\frac{\hbar}{e}U(\partial_a U^\dagger)$ is a gauge potential related to the magnetization and the curved geometry of the device. Note that the transformed Hamiltonian is still general in the sense that as of now, no decision has been taken as to which axis the local frame should rotate to. Therefore, the physics that one would like to "see" lies in this important decision, i.e. the choice of the transformation operator $U$. And in this paper, spin-orbit torque physics would become apparent in a frame rotation that rotates the $Z$ axis to the magnetization axis $e_m$ as follows

$$U\eta_m = \eta_z \quad , \quad U\sigma^m U^\dagger = \sigma^z$$

(18)

where $\eta_m$ and $\eta_z$ are respectively, the eigenstates along $e_m$ and $e_z$. And $U(\theta_m, \phi_m)$ operates in the space of the magnetic moment $m$. The frame rotation takes place in the presence of spin-orbit coupling. Upon transformation, local gauge potentials would appear in the Hamiltonian. Rewriting the Hamiltonian, one has

$$H' = \left(\frac{1}{2m_e}\right)(-i\hbar\nabla_a + eA_a + US_aU^\dagger)(-i\hbar\partial^a + eA_a + US^aU^\dagger) + G_{ab}U\sigma^a m^b U^\dagger$$

(19)

The physics of curve, spin-orbit coupling, and magnetism is now captured in the expression of a total gauge: $\mathbb{Q}_a = eA_a + US_aU^\dagger$. In fact, the rotation gauge $eA_a$ has previously been associated with a form of adiabatic spin-tansfer torque **[34]** that would not be further elaborated in this article. As our interest lies in the spin-orbit torque, we will only focus on the transformed spin-orbit gauge also known henceforth as $\mathcal{F}_a = US_aU^\dagger$. Relevant to our study are the energy terms as follows:

$$E = \psi^\dagger\left(\frac{1}{2m_e}\right)(-i\hbar\nabla_a + \mathcal{F}_a)(-i\hbar\partial^a + \mathcal{F}^a)\psi$$

(20)

Dropping the second-order derivative kinetic energy terms,

$$E_{int} = \left(\frac{-i\hbar}{2m_e}\right)\psi^\dagger[\nabla_a \mathcal{F}^a + \mathcal{F}_a \partial^a]\psi$$

(21)

Now, we will examine each energy density ($\nabla_a \mathcal{F}^a$, $\mathcal{F}_a \partial^a$) term in details:

$$\nabla_a \mathcal{F}^a \rightarrow \left(\frac{-i\hbar}{2m_e}\right)[\partial_a(\psi^\dagger \mathcal{F}^a \psi) - (\partial_a \psi^\dagger)\mathcal{F}^a \psi + \Gamma_a(\psi^\dagger \mathcal{F}^a \psi)]$$

(22)

$$\mathcal{F}_a \partial^a \rightarrow \left(\frac{-i\hbar}{2m_e}\right)\psi^\dagger \mathcal{F}_a \partial^a \psi$$

(23)

The expression $\nabla_a \mathcal{F}^a$ comprises three terms on the RHS. It is reduced to $\left(\frac{-i\hbar}{2m_e}\right)[-(\partial_a\psi^\dagger)\mathcal{F}^a\psi]$ when the first and the third term vanish as surface terms as the energy density is integrated over the entire device

$$\sqrt{G}\int[\partial_a + \Gamma_a](\psi^\dagger \mathcal{F}^a \psi)\, dV = \sqrt{G}\int(\psi^\dagger \mathcal{F}^a \psi).dS = 0$$

(24)

It suffices now to proceed with the remaining terms which combine to make physical sense in terms of $j_a$ taking on the physical meaning of current density as follows,

$$\left(\frac{-i\hbar}{2m}\right)[\mathcal{F}_a \psi^\dagger \partial^a \psi - (\partial_a \psi^\dagger)\mathcal{F}^a \psi] \leftrightarrow j_a \mathcal{F}^a$$

(25)

The infinitesimal bulk is then compressed along $e_3$, i.e. taking the operation of $q_3 \rightarrow 0$ where $(\sqrt{G})_{q_3 \rightarrow 0} = \sqrt{g}$. As the Z axis is rotated to the magnetic moment (**m**), and spin is assumed to align with **m** in an adiabatic manner and at all times (even as **m** changes spatially) as it propagates, the eigenspinor of the electron would undergo $\eta_m \rightarrow \eta_z$. Refering to the formulation **[35]** for the volume and surface integrals,

$$\int \psi^\dagger \psi\, dV = \int \psi^\dagger \psi\, dq_1\, dq_2\, dq_3\, \sqrt{G}$$

(26)

With $dS = \sqrt{g}\, dq_1 dq_2$

$$\int \psi^\dagger \psi\, dV = \int \psi^\dagger \psi\, f\, dS\, dq_3$$

(27)

From the above, $dV = f\, dS\, dq_3$. Now, writing $\int \psi^\dagger \psi\, dV = \int \chi^\dagger \chi\, dS\, dq_3$ means the wave-functions are taken to be $\psi = \frac{\chi\, \eta_z}{\sqrt{f}}$, $\psi^\dagger = \frac{\eta_z^\dagger \chi^\dagger}{\sqrt{f}}$ and $\sqrt{f} = \frac{\sqrt{G}}{\sqrt{g}}$. Note that $\chi = \chi_s \chi_n$ are separable scalar functions where $\chi_s$ is the surface wave-function, and $\chi_n$ is normal to the surface. The energy terms can now be written as follows:

$$\psi^\dagger \mathcal{F}_a \partial^a \psi = \left( \frac{\eta_z^\dagger \chi^\dagger}{f} \mathcal{F}_a (\partial^a \chi) \eta_z + \frac{\eta_z^\dagger \chi^\dagger}{\sqrt{f}} \mathcal{F}_a \chi \eta_z \left( \partial^a f^{-\frac{1}{2}} \right) \right) \tag{28}$$

$$-(\partial_a \psi^\dagger) \mathcal{F}^a \psi = \left( -\frac{(\partial^a \eta_z^\dagger \chi^\dagger)}{f} \mathcal{F}_a \chi \eta_z - \frac{\eta_z^\dagger \chi^\dagger}{\sqrt{f}} \left( \partial^a f^{-\frac{1}{2}} \right) \mathcal{F}_a \chi \eta_z \right) \tag{29}$$

Combining the above,

$$\psi^\dagger \mathcal{F}_a \partial^a \psi - (\partial_a \psi^\dagger) \mathcal{F}^a \psi = \frac{1}{f} \langle \eta_z | \mathcal{F}_a\, J^a | \eta_z \rangle \tag{30}$$

where $J^a = \chi^\dagger \partial^a \chi - (\partial_a \chi^\dagger) \chi$ is now the current density. Since $\mathcal{F}^a$ orginates from the spin-orbit coupling in a curved space, the entire term $j_a \mathcal{F}^a$ would be an interaction energy density ($E_{int}$) related to the coupling of the current flow ($j_a$) with the curved spin-orbit physics. As $\chi = \chi_s(q_1, q_2)\, \chi_n(q_3)$, the term $j_a \mathcal{F}^a$ is decoupled into

$$j_a \mathcal{F}^a = (j_1 \mathcal{F}^1 + j_2 \mathcal{F}^2) + j_3 \mathcal{F}^3 \tag{31}$$

where $j_1 \mathcal{F}^1 + j_2 \mathcal{F}^2$ is the surface current density and $j_3 \mathcal{F}^3$ the normal current density. Recall that $\mathcal{F}_a = U S_a U^\dagger$, and $S_a$ is summarized below

$$\begin{pmatrix} S_1 \\ S_2 \\ S_3 \end{pmatrix} = \frac{m_e \sqrt{G}}{\hbar} \begin{pmatrix} 0 & -\alpha^3 & \alpha^2 \\ \alpha^3 & 0 & -\alpha^1 \\ -\alpha^2 & \alpha^1 & 0 \end{pmatrix} \begin{pmatrix} \sigma^1 \\ \sigma^2 \\ \sigma^3 \end{pmatrix} \tag{32}$$

To confine the spin-orbit effect to the 2D which is prevalent in physical systems like the Rashba, 2D Dressellhaus, topological surface states, 2D graphene and silicene, one takes the limit of $q_3 \to 0$ and notes that $f \to 1$. It's worth noting that the terms $\frac{\eta_z^\dagger \chi^\dagger}{\sqrt{f}} \left( \partial^a f^{-\frac{1}{2}} \right) \mathcal{F}_a \chi \eta_z$ would vanish for $a = 1, 2$ as limit $q_3 \to 0$ is taken. The surviving term $\frac{\eta_z^\dagger \chi^\dagger}{\sqrt{f}} \left( \partial^3 f^{-\frac{1}{2}} \right) \mathcal{F}_3 \chi \eta_z$ would be expected to capture the physics of the confinement effect. But as steps have been taken to ensure Hermiticty in spin-orbit coupling Eq.(14), the confinement terms of Eq.(28) and Eq.(29) cancel one another. This is an unexpected effect for the spin-orbit torque that eliminates a curved-surface confinement effect because of the symmetrization in the current density. With the limit taken, the normal current density is discarded and one should from now on ignore the effect of $S_3$. We will proceed to the spin-orbit constant which is given by $\alpha^n = \alpha' E^n$. We will now let $E^1$ and $E^2$ approach zero, so that only $\alpha^3$ is retained. The matrix is reduced back the 2D formalism like in the beginning of the article,

$$\begin{pmatrix} S_1 \\ S_2 \end{pmatrix} = \frac{m_e \sqrt{g}}{\hbar} \begin{pmatrix} 0 & -\alpha^3 \\ \alpha^3 & 0 \end{pmatrix} \begin{pmatrix} \sigma^1 \\ \sigma^2 \end{pmatrix} \tag{33}$$

It's worth noting that there's a caveat in the limit taking, i.e. we have let $q_3 \to 0$ first so that the surface confinement term $\frac{\eta_z^\dagger \chi^\dagger}{\sqrt{f}} \left( \partial^3 f^{-\frac{1}{2}} \right) \mathcal{F}_3 \chi \eta_z$ survived although it would vanish later because of symmetrization. Had the limit of $\alpha^1 \to 0, \alpha^2 \to 0$ been taken first, the confinement term would not have shown up. Therefore, the general spin-orbit gauge potentials can now be expressed as follows:

$$\mathcal{F}_1 = -U\left(\frac{\alpha^3 m_e}{\hbar}\sqrt{g}\sigma^2\right)U^\dagger, \quad \mathcal{F}_2 = U\left(\frac{\alpha^3 m_e}{\hbar}\sqrt{g}\sigma^1\right)U^\dagger$$

(34)

Finally, the energy density is

$$E_{int} = j_a \mathcal{F}^a = \left(\frac{\alpha^3 m}{e\hbar}\sqrt{g}\right)\left[-j^1\, U\sigma^2 U^\dagger + j^2\, U\sigma^1 U^\dagger\right]$$

(35)

The energy density in magnetic space is an important development for the study of magnetic torque and flipping via the concept of the anisotropy field. We consider a potential in the magnetic space to arise from the energy density $E_{int}$. The potential is essentially the effective anisotropy field as it is commonly understood in magnetic physics:

$$\mu\, \boldsymbol{H} = \frac{\partial E_{int}}{\partial \boldsymbol{m}} = \frac{\partial}{\partial \boldsymbol{m}}\left(\frac{j^a \mathcal{F}_a}{e}\right)$$

(36)

The explicit expression of $\mu\, \boldsymbol{H}$ is therefore, given as

$$\mu\, \boldsymbol{H} = \left(\frac{\alpha^3 m_e}{\hbar}\sqrt{g}\right)\left[\frac{-j^1}{e}\left(\frac{\partial}{\partial \boldsymbol{m}} U\sigma^2 U^\dagger\right) + \frac{j^2}{e}\left(\frac{\partial}{\partial \boldsymbol{m}} U\sigma^1 U^\dagger\right)\right]$$

(37)

The magnetic moment ($\boldsymbol{m}$) of the ferromagnetic material is superimposed on the curved surface as shown in Fig.3. below. The curved surface can be accessed at every point in space by $\boldsymbol{R} = \boldsymbol{r}(q_1, q_2) + q_3 \boldsymbol{e_3}$ that charcaterizes the surface on which $\boldsymbol{m}$ is located. Vector $\boldsymbol{m}$ will also be indexed by $(m^1, m^2)$ to reflect its components in the general coordinates of the real space. In that way, the magnetic components will also track the structure of the space in which it is superimposed. The actual orientation of $\boldsymbol{m}$ is described by $(\theta^m, \phi^m, m)$ with respect to the Cartesian coordinates, where superscripts of $(\theta^m, \phi^m)$ merely show that these are coordinates for the magnetization.

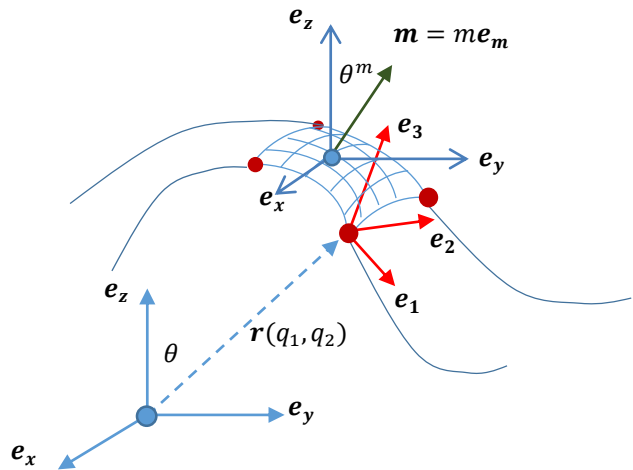

**FIG.3** A 2D curve is embeded in a 3D space and characterized by parameters $(q_1, q_2, q_3)$. The magnetization $\boldsymbol{m}$ is superimposed on the curved surface.

We will now look into the spin physics and focus our attention on the Pauli matrices. Originally expressed in the general corodinates, the Pauli matrices are now re-expressed in the Cartesian coordinates in which the adiabatic physics will be introduced at a later stage. In the physics of forces, and magnetism, the important physical quantities are: magnetic moment ($\boldsymbol{m}$), Pauli matrix ($\sigma$), current density ($j$). There are frame of references under which the physical quantities most relevant to our studies are described, e.g. the Cartesian $(x, y, z)$, and the geneal coordinates $(q_1, q_2, q_3)$. Physical quantities and their dimensions are expressed in both frames. For example, $\boldsymbol{m} = m^1 \boldsymbol{e_1} + m^2 \boldsymbol{e_2} = m^v \partial_v^a \boldsymbol{e_a}$, implies that a physical quantity remains unchanged when expressed under any frame of reference. Explicitly, it will look as follows

$$\boldsymbol{m} = m^x \boldsymbol{e_x} + m^y \boldsymbol{e_y} + m^z \boldsymbol{e_z} = m^1 \boldsymbol{e_1} + m^2 \boldsymbol{e_2} + m^3 \boldsymbol{e_3}$$
(38)

The two are connected by the transformaton as follows

$$\boldsymbol{m} = (m^1 \partial_1^x + m^2 \partial_2^x + m^3 \partial_3^x)\boldsymbol{e_x} + (m^1 \partial_1^y + m^2 \partial_2^y + m^3 \partial_3^y)\boldsymbol{e_y} + (m^1 \partial_1^z + m^2 \partial_2^z + m^3 \partial_3^z)\boldsymbol{e_z}$$
(39)

The above provides an illustration of coordinate transformation. In the actual context, the same principle is applied to the Pauli matrices to reappear in the Cartesian frame. As a result, the effective anisotropy field would now be,

$$\mu\,\boldsymbol{\delta H} = \left(\frac{\alpha^3 m_e}{\hbar m}\sqrt{g}\right)\left[-\frac{j^1}{e}\frac{\partial}{\partial \boldsymbol{n}}\left(\partial_v^2 U \sigma^v U^\dagger\right) + \frac{j^2}{e}\frac{\partial}{\partial \boldsymbol{n}}\left(\partial_v^1 U \sigma^v U^\dagger\right)\right]$$
(40)

where $v$ runs over the Cartesian $(x, y, z)$. In the real space, we note once again that the Pauli matrices in the general coordinates have been re-expressed in the Cartesian frame whereby the curve effects would be reflected in the quantities of $\partial_v^2$ and $\partial_v^1$. On the other hand, the unitary operator $U$ is now applied to rotate the magnetic axis $(\boldsymbol{e_m})$ to the $(x, y, z)$ frame, thereby providing the $\boldsymbol{m}$ an indirect link in terms of orientation to the general coordinates. It is now possible to associate the Pauli matices with $\boldsymbol{m}$ and perform the operation of $\frac{\partial}{\partial \boldsymbol{n}}$ for the anisotropy field as required in Eq.(38). In the following, we provide an explicit demonstration of the $U$ operation. Recall that $U$ is parameterized by $(\theta^m, \phi^m)$. Refer to Fig.4 below and observe that $U$ rotates the eigenstate along $\boldsymbol{e_m}$ to the Z axis about axis $\boldsymbol{e_{m2}} = -\boldsymbol{e_y}\cos\phi^m + \boldsymbol{e_x}\sin\phi^m$.

$$U = \begin{pmatrix} \cos\frac{\theta^m}{2} & \sin\frac{\theta^m}{2} e^{-i\phi^m} \\ -\sin\frac{\theta^m}{2} e^{i\phi^m} & \cos\frac{\theta^m}{2} \end{pmatrix}$$
(41)

Rotation of the magnetic axis $(\boldsymbol{e_m})$ to the $(x, y, z)$ frame is given by

$$U\sigma U^\dagger = \sigma^z \boldsymbol{e_m} + \sigma^a \boldsymbol{e_{m1}} + \sigma^{m2} \boldsymbol{e_{m2}}$$
(42)

Refer to Fig.4 for the directions implied by the superscripts and subscripts in the equation above. Note that the $\boldsymbol{m}$ physical quantities of $(\boldsymbol{e_m}, \boldsymbol{e_{m1}}, \boldsymbol{e_{m2}})$ and $(\sigma^a, \sigma^{m2})$ can all be related to the $(x, y, z)$ frame as follows

$$\boldsymbol{e_m} = \boldsymbol{e_x}\sin\theta^m \cos\phi^m + \boldsymbol{e_y}\sin\theta^m \sin\phi^m + \boldsymbol{e_z}\cos\theta^m$$
$$= n^x \boldsymbol{e_x} + n^y \boldsymbol{e_y} + n^z \boldsymbol{e_z}$$

$$e_{m1} = -e_y \cos\theta^m \sin\phi^m - e_x \cos\theta^m \cos\phi^m + e_z \sin\theta^m \quad (43)$$

$$e_{m2} = -e_y \cos\phi^m + e_x \sin\phi^m \quad (44)$$

$$\sigma^a = -\sigma^x \cos\phi^m - \sigma^y \sin\phi^m \quad (45)$$

$$\sigma^{m2} = \sigma^x \sin\phi^m - \sigma^y \cos\phi^m \quad (46)$$

$$(47)$$

The following is a schematic showing the magnetic moment vectors in the Cartesian frame.

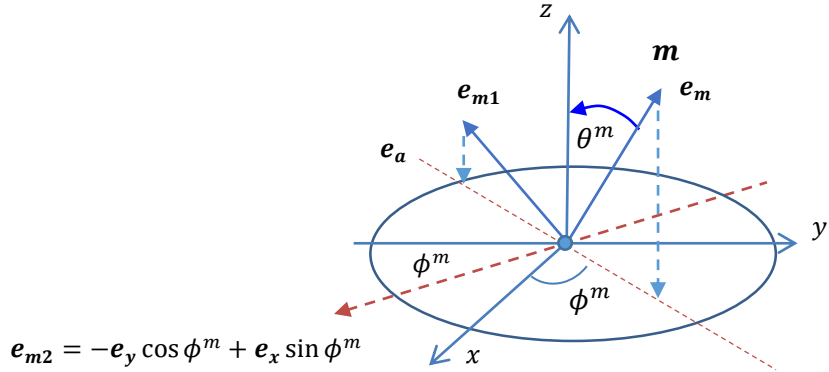

**FIG.4.** Spin rotation about $e_{m2}$ moves the Z axis to the magnetic moment axis $e_m$, and the $e_a$ axis to $e_{m1}$ or vice versa.

Following the mark of $U\sigma^v U^\dagger$ in Eq.(40) where $v$ runs over $(x,y,z)$, it is handy to write down the following for the sake of easy inspection.

$$U\boldsymbol{\sigma} U^\dagger = U\sigma^x U^\dagger e_x + U\sigma^y U^\dagger e_y + U\sigma^z U^\dagger e_z$$

$$(48)$$

Using Eqs. (42) to (47), the process of linking the original magnetic axis through unitary rotation to the $(x,y,z)$ frame is complete, where

$$U\sigma^x U^\dagger = \sin\theta^m \cos\phi^m\ \sigma^z + f(\sigma^a, \sigma^{m2})$$

$$(49)$$

$$U\sigma^y U^\dagger = \sin\theta^m \sin\phi^m\ \sigma^z + g(\sigma^a, \sigma^{m2})$$

$$(50)$$

$$U\sigma^z U^\dagger = \cos\theta^m\ \sigma^z + h(\sigma^a)$$

$$(51)$$

Note that $f, g, h$ are linear combinations of $\sigma^a, \sigma^{m2}$ and these terms are non-issues as they would vanish later. At this point, an important physical step is taken, i.e. the adiabatic approximation in which it is assumed that electron spin aligns along the new Z axis that has been rotated to $e_m$. Under the adiabatic spin alignment whereby the following is resulted: $\langle \eta_z | \sigma^x | \eta_z \rangle = 0$, $\langle \eta_z | \sigma^y | \eta_z \rangle = 0$, $\langle \eta_z | \sigma^z | \eta_z \rangle = 1$, functions $f, g, h$ vanish while the following remains

$$\langle \eta_z | U\sigma^x U^\dagger | \eta_z \rangle = n^x, \qquad \langle \eta_z | U\sigma^y U^\dagger | \eta_z \rangle = n^y, \qquad \langle \eta_z | U\sigma^z U^\dagger | \eta_z \rangle = n^z$$

(52)

It thus follows that

$$\langle \eta_z | \mu \, \boldsymbol{\delta H} | \eta_z \rangle = \left( \frac{\alpha^3 m}{\hbar M_s} \sqrt{g} \right) \left[ -\frac{j^1}{e} \frac{\partial}{\partial \boldsymbol{n}} (\partial_v^2 n^v) + \frac{j^2}{e} \frac{\partial}{\partial \boldsymbol{n}} (\partial_v^1 n^v) \right]$$

$$= \left( \frac{\alpha^3 m}{e\hbar M_s} \sqrt{g} \right) [-j^1 (\partial_v^2 \boldsymbol{e}^v) + j^2 (\partial_v^1 \boldsymbol{e}^v)]$$

(53)

Note in the above that $n^v$ carries a contravariant index. But, $n_a$ of $\frac{\partial}{\partial \boldsymbol{n}} = \frac{\partial}{\partial n_a} \boldsymbol{e}_a$ is covariant. As a result,

$$\langle \eta_z | \mu \, \boldsymbol{\delta H} | \eta_z \rangle = \left( \frac{\alpha^3 m}{e\hbar M_s} \sqrt{g} \right) [-j^1 \boldsymbol{e}^2 + j^2 \boldsymbol{e}^1]$$

(54)

*Equation (54) is the general form of what we called the spin-orbit anisotropy field. This is another important result of this paper.* Note that $(j^1, j^2)$ and $(\boldsymbol{e}^1, \boldsymbol{e}^2)$ can be found by transforming their Cartesian counterparts to the desired coordinates in a contravariant manner. Such formulation is important because as device becomes smaller, the physics of local geometries on spin-orbit effect and its torque shall not be neglected. The general formulation allows spin and magnetic dynamic to be studied on local curved surfaces. With the spin-orbit anisotropy field derived, the spin-orbit torque is a straightforward cross product with the $\boldsymbol{m}$

$$\boldsymbol{\tau} = -\boldsymbol{m} \times \langle z | \mu \, \boldsymbol{\delta H} | z \rangle$$

(55)

As the formulation is provided in general coordinates, the spin-orbit effective field can be derived for any surfaces to be characterized by $\boldsymbol{R} = \boldsymbol{r}'(q_1, q_2) + q_3 \boldsymbol{e}_3$. For illustration, we will now take the examples of the spherical, cylindrical and flat Cartesian surfaces. In the *spherical system,* the coordinates are $\boldsymbol{R} = \boldsymbol{r}'(\theta, \phi) + \delta r \boldsymbol{e}_r$ and

$$\langle z | \mu \, \boldsymbol{\delta H} | z \rangle = \left( \frac{\alpha^3 m}{e\hbar M_s} \sqrt{g} \right) \left( -\frac{\partial \theta}{\partial v} \frac{\partial \phi}{\partial v'} + \frac{\partial \phi}{\partial v} \frac{\partial \theta}{\partial v'} \right) j^v \boldsymbol{e}^{v'}$$

(56)

The current carrying carriers are constrained to the surface, i.e. $\delta r = 0$. The spherical surface is then accessed by $\boldsymbol{r}' = (r \sin\theta \cos\phi, r \sin\theta \sin\phi, r \cos\theta)$

$$\begin{pmatrix} \boldsymbol{e}_x \\ \boldsymbol{e}_y \\ \boldsymbol{e}_z \end{pmatrix} = \begin{pmatrix} \frac{\partial \theta}{\partial x} & \frac{\partial \phi}{\partial x} & \frac{\partial r}{\partial x} \\ \frac{\partial \theta}{\partial y} & \frac{\partial \phi}{\partial y} & \frac{\partial r}{\partial y} \\ \frac{\partial \theta}{\partial z} & \frac{\partial \phi}{\partial z} & \frac{\partial r}{\partial z} \end{pmatrix} \begin{pmatrix} \boldsymbol{e}_\theta \\ \boldsymbol{e}_\phi \\ \boldsymbol{e}_R \end{pmatrix} = \begin{pmatrix} \frac{\cos\phi \cos\theta}{r} & \frac{\sin\phi}{-r \sin\theta} & \sin\theta \cos\phi \\ \frac{\sin\phi \cos\theta}{r} & \frac{\cos\phi}{r \sin\theta} & \sin\theta \sin\phi \\ \frac{-\sin\theta}{r} & 0 & \cos\theta \end{pmatrix} \begin{pmatrix} \boldsymbol{e}_\theta \\ \boldsymbol{e}_\phi \\ \boldsymbol{e}_r \end{pmatrix}$$

(57)

Note that $v, v'$ run over $(x, y, z)$ and summation is only non-vanishing for $v \neq v'$. Meanwhile, $g = \sqrt{(g_{\theta\theta} g_{\phi\phi} - g_{\theta\phi} g_{\phi\theta}) g_{rr}} = r^2 \sin\theta$.

$$\langle z | \mu \, \boldsymbol{\delta H} | z \rangle = \left( \frac{\alpha^3 m}{e\hbar M_s} \sqrt{g} \right) \left( \frac{1}{r^2 \tan\theta} (-j_x \boldsymbol{e}_y + j_y \boldsymbol{e}_x) + \frac{\sin\phi}{r^2} (j_x \boldsymbol{e}_z - j_z \boldsymbol{e}_x) + \frac{\cos\phi}{r^2} (j_z \boldsymbol{e}_y - j_y \boldsymbol{e}_z) \right)$$

$$= \left(\frac{\alpha^3 m}{e\hbar M_s}\right)\left(\cos\theta\left(-j_x\boldsymbol{e_y} + j_y\boldsymbol{e_x}\right) + \sin\theta\sin\phi\left(j_x\boldsymbol{e_z} - j_z\boldsymbol{e_x}\right) + \sin\theta\cos\phi\left(j_z\boldsymbol{e_y} - j_y\boldsymbol{e_z}\right)\right)$$

(58)

With the above, the spin-orbit anisotropy field can be estimated over the entire spherical surface provided $(j_x, j_y, j_z)$ is computed or measured over the surface. Likewise, with $(m_x, m_y, m_z)$ computed or measured over the surface, the spin-orbit torque can be determined at every point of the surface. Figure 5 illustrates how current flows into and out of the nanoscale structures. Devices' central region can be fabricated to these structures and one can regard the current to flow from source to drain.

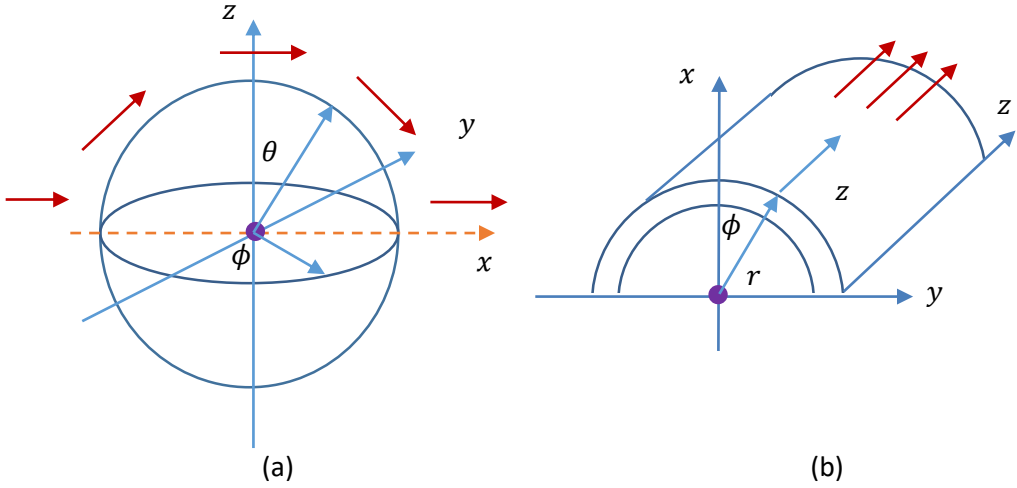

(a)              (b)

**FIG.5** (a) A spherical surface across which current flows from left to right as indicated by arrows. (b) A cylindrical surface across which current flows along the z direction as shown by the arrow.

In the *cylindrical system*, the coordinates are $\boldsymbol{R} = \boldsymbol{r}'(\phi, z) + \delta r \boldsymbol{e_r}$. The current carrying carriers are constrained to the surface, i.e. $\delta r = 0$. The cylindrical surface is then accessed by $\boldsymbol{r}' = (r\cos\phi, r\sin\phi, z)$.

$$\begin{pmatrix}\boldsymbol{e_x}\\ \boldsymbol{e_y}\\ \boldsymbol{e_z}\end{pmatrix} = \begin{pmatrix}\frac{\partial\phi}{\partial x} & \frac{\partial z}{\partial x} & \frac{\partial r}{\partial x}\\ \frac{\partial\phi}{\partial y} & \frac{\partial z}{\partial y} & \frac{\partial r}{\partial y}\\ \frac{\partial\phi}{\partial z} & \frac{\partial z}{\partial z} & \frac{\partial r}{\partial z}\end{pmatrix}\begin{pmatrix}\boldsymbol{e_\phi}\\ \boldsymbol{e_z}\\ \boldsymbol{e_r}\end{pmatrix} = \begin{pmatrix}-\frac{\sin\phi}{r} & 0 & \cos\phi\\ +\frac{\cos\phi}{r} & 0 & \sin\phi\\ 0 & 1 & 0\end{pmatrix}\begin{pmatrix}\boldsymbol{e_\phi}\\ \boldsymbol{e_z}\\ \boldsymbol{e_r}\end{pmatrix}$$

(59)

It follows that

$$\langle z|\mu\,\boldsymbol{\delta H}|z\rangle = \left(\frac{\alpha^3 m}{e\hbar M_s}\sqrt{g}\right)\left(\frac{\sin\phi}{r}(j_x\boldsymbol{e_z} - j_z\boldsymbol{e_x}) + \frac{\cos\phi}{r}(j_z\boldsymbol{e_y} - j_y\boldsymbol{e_z})\right)$$

(60)

Note that $v, v'$ run over $(x, y, z)$ and summation is only non-vanishing for $v \neq v'$. In the meantime, $g = \sqrt{(g_{\phi\phi}g_{zz} - g_{\phi z}g_{z\phi})g_{rr}} = r$. Thus,

$$\langle z|\mu\,\boldsymbol{\delta H}|z\rangle = \left(\frac{\alpha^3 m}{e\hbar M_s}\right)\left(\sin\phi\,(j_x\boldsymbol{e_z} - j_z\boldsymbol{e_x}) + \cos\phi\,(j_z\boldsymbol{e_y} - j_y\boldsymbol{e_z})\right)$$

(61)

Last, in the *Cartesian system*, physical quantities are re-expressed in the *x-y* basis. The results are

$$\langle z|\mu\, \boldsymbol{\delta H}|z\rangle = \left(\frac{\alpha^z m}{e\hbar M_s}\right)\left[-j^x \boldsymbol{e_y} + j^y \boldsymbol{e_x}\right]$$

(62)

The spin-orbit anisotropy field above when substituted in the standard expression for spin torque in a flat surface device, would lead to the to the well-known, experimentally proven **[20, 21]** spin-orbit torque, also known to the experimental community as the field-like spin-orbit torque. This is, however, not the same as another form of spin-orbit torque aka the anti-damping spin-orbit torque **[22]**. In other words, what we have derived is the general formulation for the field-like spin-orbit spin torque.

## CONCLUSION

We have provided a general formulation of the spin-orbit coupling on a curved surface in Eq.(5) with $\alpha^3$ and hence $(S_1, S_2)$ properly defined and understood. We had then considered a curved ferromagnetic and oxide heterostructure to give rise to a spin-orbit torque. A proper choice of the transformation operator $U$ that rotates the Z axis to the magnetization axis $\boldsymbol{e_m}$ leads to the derivation of the spin-orbit anisotropy field and hence the spin-orbit torque. We note that an unexpected effect that arises in the symmetriation of the current density actually eliminates the curved-surface confinement. Finally, with the adiabatic approximation, we completed the general formluation of the spin-orbit anisotropy field in Eq.(54) and hence the spin-orbit torque that can be computed over the entire surface of devices of any shape. We provided examples in spherical, cylindrical and the Cartesian surfaces

**Acknowledgments**

We would like to thank the National Science and Technology Council of Taiwan for supporting this work under Grant No. 110-2112-M-034-001-MY3.

**ORCID iD**

Seng Ghee Tan ・ https://orcid.org/0000-0002-3233-9207